\documentclass[aps,showpacs,nofootinbib,twocolumn]{revtex4}

\usepackage{latexsym}
\usepackage{epsfig}
\usepackage{amsmath}
\usepackage{amssymb}
\usepackage{graphicx}

\newcommand{\lp}{\left(}
\newcommand{\rp}{\right)}

\newcommand{\ba}{\begin{eqnarray}}
\newcommand{\ea}{\end{eqnarray}}
\newcommand{\be}{\begin{equation}}
\newcommand{\ee}{\end{equation}}

\newcommand{\ka}{\kappa}

\newcommand{\R}{\mathcal{R}}

\begin{document}

\title{Einstein static Universe in hybrid metric-Palatini gravity}

\author{Christian G. B\"ohmer$^1$}\email{c.boehmer@ucl.ac.uk}
\author{Francisco S.N.~Lobo$^{2}$}\email{flobo@cii.fc.ul.pt}
\author{Nicola Tamanini$^1$}\email{n.tamanini.11@ucl.ac.uk}

\affiliation {$^1$Department of Mathematics and Institute of Origins, University College London, Gower Street, London, WC1E 6BT, UK}
\affiliation{$^2$Centro de Astronomia e Astrof\'{\i}sica da Universidade de Lisboa, Campo Grande, Ed. C8 1749-016 Lisboa, Portugal}

\date{\today}

\begin{abstract}

Hybrid metric-Palatini gravity is a recent and novel approach to modified theories of gravity, which consists of adding to the metric Einstein-Hilbert Lagrangian an $f(\R)$ term constructed \`{a} la Palatini. It was shown that the theory passes local tests even if the scalar field is very light, and thus implies the existence of a long-range scalar field, which is able to modify the dynamics in galactic and cosmological scales, but leaves the Solar System unaffected. In this work, motivated by the possibility that the Universe may have started out in an asymptotically Einstein static state in the inflationary universe context, we analyse the stability of the Einstein static Universe by considering linear perturbations in the respective dynamically equivalent scalar-tensor representation of hybrid metric-Palatini gravity. Considering linear homogeneous and inhomogeneous perturbations, the stability regions of the Einstein static universe are parametrized by the first and second derivatives of the scalar potential, and it is explicitly shown that a large class of stable solutions exists in the respective parameter space,  in the context of hybrid metric-Palatini gravity.

\end{abstract}


\maketitle

\section{Introduction}

The standard model of cosmology is remarkably successful in accounting for the observed features of the Universe and during the last two decades has evolved from being mainly a theoretical area of Physics to become a field supported by observational data of high-precision. Recent experiments call upon state of the art technology in Astronomy and in Astrophysics to provide detailed information about the contents and history of the Universe. This has led to the measuring of the parameters that describe our Universe with increasing precision. In this context, the Nobel Prize of Physics 2011 was attributed for the discovery of the accelerating expansion of the Universe through observations of distant supernovae \cite{cosmic_acc}. In fact, after fifteen years of extensive research, we still lack a fundamental understanding of this late-time cosmic acceleration, and its resolution will be one of the major objectives in cosmology during the next decade with several surveys and experiments to address the nature of the cosmic speed-up. Recently, the released Planck data of the 2.7 full sky survey \cite{Planck} have also shown a number of intriguing features, which will require a change in our standard view of the Universe. 

The standard model of cosmology has favoured a missing stress-energy component, in particular, the dark energy models \cite{Copeland:2006wr}, although one may also explore the alternative viewpoint, namely, through a modified gravity approach, in particular, using generalizations of the Einstein-Hilbert Lagrangian \cite{modgrav1,modgrav2,modgrav3,Boehmer:2013ss}. In fact, it is fundamental to understand how one may differentiate these modified theories of gravity from dark energy models. Note that generalizations of the action functional can be approached in several ways. For instance, prescriptions consist in replacing the linear scalar curvature term in the Einstein-Hilbert action by a function of the scalar curvature, $f(R)$ gravity \cite{modgrav1}, or by more general scalar invariants of the theory \cite{modgrav2}, or still considering a coupling between curvature and matter \cite{modgrav3}, a class of theories often termed higher-order gravity theories. Different approaches have been analysed in the literature: The metric formalism \cite{modgrav1}, which consists in varying the action with respect to $g^{\mu\nu}$; the Palatini formalism~\cite{Palatini,Sotiriou:2006qn}, where the metric and the connections are treated as separate variables; and the metric-affine formalism, where the matter part of the action now depends and is varied with respect to the connection~\cite{Sotiriou:2006qn}.

A novel approach to modified theories of gravity was recently proposed, consisting of adding to the metric Einstein-Hilbert Lagrangian an $f(\R)$ term constructed \`{a} la Palatini~\cite{Harko:2011nh}. It was shown that using the respective dynamically equivalent scalar-tensor representation, the theory can pass Solar System observational constraints even if the scalar field is very light. This implies the existence of a long-range scalar field, which is able to modify the cosmological \cite{Capozziello:2012ny} and galactic dynamics~\cite{Capozziello:2012qt}, but leaves the Solar System unaffected. Static and spherically symmetric solutions, in particular, wormhole solutions, were analysed in~\cite{Capozziello:2012hr}.
A generalization of the hybrid metric-Palatini gravitational theory was also presented~\cite{Tamanini:2013ltp}, where the gravitational action is taken to depend on a general function of both the metric and Palatini curvature scalars. The dynamical equivalence with a non-minimally coupled bi-scalar field gravitational theory was proved and the evolution of cosmological solutions was studied using dynamical systems techniques. In~\cite{koivTam} the ghost analysis for these theories was recently performed showing that this biscalar generalization always presents either ghosts or tachyons. In fact, new types of second, fourth and sixth order derivative gravity theories were investigated and the hybrid metric-Palatini theory was singled out as a viable class of ``hybrid'' extensions of General Relativity.

The cosmological applications of this hybrid metric-Palatini gravitational theory were also further explored~\cite{Capozziello:2012ny}. More specifically, cosmological solutions coming from its scalar-tensor representation of hybrid metric-Palatini gravity were found and criteria to obtain the cosmic acceleration were discussed. Furthermore, several classes of dynamical cosmological solutions, depending on the functional form of the effective scalar field potential, describing both accelerating and decelerating Universes were explicitly obtained. The cosmological perturbation equations were also derived and applied to uncover the nature of the propagating scalar degree of freedom and the signatures these models predict in the large-scale structure.

Relative to the galactic dynamics, the generalized virial theorem and certain astrophysical applications in the scalar-tensor representation of the hybrid metric-Palatini gravity were explored in~\cite{Capozziello:2012qt}. More specifically, taking into account the relativistic collisionless Boltzmann equation, it was shown that the supplementary geometric terms in the gravitational field equations provide an effective contribution to the gravitational potential energy. It was also shown that the total virial mass is proportional to the effective mass associated with the new terms generated by the effective scalar field, and the baryonic mass. This fact demonstrates that the geometric origin in the generalized virial theorem may account for the well-known virial theorem mass discrepancy in clusters of galaxies. The possibility that the behavior of the rotational velocities of test particles gravitating around galaxies can be explained within the framework of the hybrid metric-Palatini gravitational theory, was also recently  explored \cite{Capozziello:2013yha}. Thus, hybrid metric-Palatini gravity provides an effective alternative to the dark matter paradigm of present day cosmology and astrophysics.

In this work, in the cosmological context, motivated by the fact that the Einstein static Universe has always been of great interest in various gravitational theories, and inspired by the possibility that the universe may have started out in an asymptotically Einstein static state in the inflationary universe context~\cite{Ellis:2002we}, we analyse the Einstein static Universe and its stability in the hybrid metric-Palatini gravitational theory. In fact, the Einstein static universe has been analysed in a plethora of contexts, such as: in General Relativity with a non-constant pressure \cite{ESa}; in brane world models~\cite{Gergely:2001tn}; in Einstein-Cartan theory~\cite{Boehmer:2003iv}; in loop quantum cosmology~\cite{Mulryne:2005ef}; in $f(R)$ modified theories of gravity~\cite{Boehmer:2007tr,Goswami:2008fs,Seahra:2009ft}; in modified Gauss-Bonnet gravity~\cite{Bohmer:2009fc}; in IR modified Ho\v{r}ava gravity~\cite{Boehmer:2009yz}; and in massive gravity~\cite{Parisi:2012cg}.

This present paper is outlined in the following manner: In Sec.~\ref{sect:2}, we briefly review the formalism of the hybrid metric-Palatini gravitational theory, in particular the action and field equations, and in Sec.~\ref{sect:3} present the respective modified Friedman equations. In Sec.~\ref{sect:4}, we consider linear homogeneous and inhomogeneous perturbations in the context of the Einstein static Universe in the above-mentioned hybrid 
metric-Palatini theory, and analyze the respective stability regions. In Sec.~\ref{sec:concl}, we summarize and discuss our results.

\section{Hybrid metric-Palatini gravity: Formalism}\label{sect:2}

The action for the hybrid metric-Palatini gravity  is
\begin{equation} \label{eq:S_hybrid}
S=\frac{1}{2\kappa^2}\int d^4 x \sqrt{-g} \left[ R + f(\R)\right] +
S_m,
\end{equation}
where $\kappa^2\equiv 8\pi G$ and we set $c=1$. $S_m$ is the matter action, $R$ is
the metric Einstein-Hilbert term, $\R  \equiv g^{\mu\nu}\R_{\mu\nu} $ is
the Palatini curvature. $\R_{\mu\nu}$ is defined in terms of
an independent torsion-less connection $\hat{\Gamma}^\alpha_{\mu\nu}$  as
\begin{equation}
\R_{\mu\nu} \equiv \hat{\Gamma}^\alpha_{\mu\nu ,\alpha} -
\hat{\Gamma}^\alpha_{\mu\alpha , \nu} +
\hat{\Gamma}^\alpha_{\alpha\lambda}\hat{\Gamma}^\lambda_{\mu\nu}
-\hat{\Gamma}^\alpha_{\mu\lambda}\hat{\Gamma}^\lambda_{\alpha\nu}\,.
\end{equation}

Varying the action (\ref{eq:S_hybrid}) with respect to the metric,
one obtains the following gravitational field equations  
\be
\label{efe} G_{\mu\nu} +
F(\R)\R_{\mu\nu}-\frac{1}{2}f(\R)g_{\mu\nu} = \ka^2 T_{\mu\nu}\,,
\ee
where the matter energy-momentum tensor is defined as 
\begin{equation}
T_{\mu\nu}
\equiv  \frac{2}{\sqrt{-g}} \frac{\delta
(\sqrt{-g}\mathcal{L}_m)}{\delta g^{\mu\nu}}
\,.
\end{equation}
The independent connection is compatible with the metric
$F(\R)g_{\mu\nu}$, which is conformal to $g_{\mu\nu}$, and where the conformal
factor is given by $F(\R) \equiv df(\R)/d\R$. The latter considerations imply that
\ba
\label{ricci} \R_{\mu\nu} & = & R_{\mu\nu} +
\frac{3}{2}\frac{1}{F^2(\R)}F(\R)_{,\mu}F(\R)_{,\nu}
    \nonumber \\
 && - \frac{1}{F(\R)}\nabla_\mu F(\R)_{,\nu} -
\frac{1}{2}\frac{1}{F(\R)}g_{\mu\nu}\Box F(\R)\,. \ea
Note that $\R$ can be obtained from the trace of the field equations (\ref{efe}), which yields 
\be 
\label{trace} F(\R)\R -2f(\R) - R = \ka^2 T \,.
\ee

Introducing an auxiliary field, the hybrid metric-Palatini action (\ref{eq:S_hybrid}) can be turned into a scalar-tensor theory given by the following action (we refer the reader to \cite{Harko:2011nh} for more details)
\begin{equation} \label{eq:S_scalar1}
S= \frac{1}{2\kappa^2}\int d^4 x \sqrt{-g} \left[ R + \phi\R-V(\phi)\right] +S_m \ .
\end{equation}
Varying this action with respect to the metric, the scalar $\phi$ and the connection yields the following field equations
\begin{eqnarray}
R_{\mu\nu}+\phi \R_{\mu\nu}-\frac{1}{2}\left(R+\phi\R-V\right)g_{\mu\nu}&=&\kappa^2 T_{\mu\nu} \,,
\label{eq:var-gab}\\
\R-V_\phi&=&0 \,, \label{eq:var-phi}\\
\hat{\nabla}_\alpha\left(\sqrt{-g}\phi g^{\mu\nu}\right)&=&0 \,, \label{eq:connection}\
\end{eqnarray}
respectively. Note that the solution of Eq.~(\ref{eq:connection}) implies that the independent connection is the Levi-Civita connection of
a metric $h_{\mu\nu}=\phi g_{\mu\nu}$. Thus we are dealing with a bi-metric theory and $\R_{\mu\nu}$ and $R_{\mu\nu}$ are related by
\begin{equation} \label{eq:conformal_Rmn}
\R_{\mu\nu}=R_{\mu\nu}+\frac{3}{2\phi^2}\partial_\mu \phi \partial_\nu \phi-\frac{1}{\phi}\left(\nabla_\mu
\nabla_\nu \phi+\frac{1}{2}g_{\mu\nu}\Box\phi\right) \ ,
\end{equation}
and consequently
\begin{equation} \label{eq:conformal_R}
\R=R+\frac{3}{2\phi^2}\partial_\mu \phi \partial^\mu \phi-\frac{3}{\phi}\Box \phi,
\end{equation}
which can be used in the action (\ref{eq:S_scalar1}) to eliminate the independent connection and obtain the following scalar-tensor representation \cite{Harko:2011nh}
\begin{eqnarray} \label{eq:S_scalar2}
S &=& \frac{1}{2\kappa^2}\int d^4 x \sqrt{-g} \left[ (1+\phi)R +\frac{3}{2\phi}\partial_\mu \phi \partial^\mu \phi
-V(\phi)\right]
\nonumber \\
&& +S_m  .
\end{eqnarray}
It is important to note that this action differs fundamentally from the $w=-3/2$ Brans-Dicke theory due to the 
coupling of the scalar to the curvature. The trace of Eq.~(\ref{eq:var-gab}) yields $-R-\phi\R+2V=\kappa^2T$, and 
using Eq.~(\ref{eq:var-phi}), takes the following useful form
\begin{equation}\label{eq:phi(X)}
2V-\phi V_\phi=\kappa^2T+R \ .
\end{equation}

Now substituting Eq.~(\ref{eq:var-phi}) and Eq.~(\ref{eq:conformal_Rmn}) in Eq.~(\ref{eq:var-gab}), the metric field equation can be written as an effective Einstein field equation, i.e.,  $G_{\mu\nu}=\kappa^2 T^{\rm eff}_{\mu\nu}$, where the effective stress-energy tensor is given by
\begin{eqnarray}
T^{\rm eff}_{\mu\nu}&=&\frac{1}{1+\phi} \Big\{ T_{\mu\nu}
 - \frac{1}{\kappa^2} \Big[ \frac{1}{2}g_{\mu\nu}\left(V+2\Box\phi\right)+
     \nonumber \\
&& \nabla_\mu\nabla_\nu\phi-\frac{3}{2\phi}\partial_\mu \phi
\;\partial_\nu \phi + \frac{3}{4\phi}g_{\mu\nu}(\partial \phi)^2 \Big] \  \Big\}  \label{effSET} .
\end{eqnarray}

The scalar field is governed by the second-order evolution equation (we refer the reader to \cite{Harko:2011nh} for more details)
\begin{equation}
  \label{eq:evol-phi}
  -\Box\phi+\frac{1}{2\phi}\partial_\mu \phi \partial^\mu
  \phi+\frac{\phi[2V-(1+\phi)V_\phi]} {3}=\frac{\phi\kappa^2}{3}T\,,
\end{equation}
which is an effective Klein-Gordon equation. Equation~(\ref{eq:evol-phi}) shows that the scalar field is dynamical, contrary to the Palatini case where $w=-3/2$. Thus, the theory is not affected by the microscopic instabilities that arise in Palatini models with infrared corrections.

\section{Einstein static Universe and modified Friedmann equations}\label{sect:3}

\subsection{Effective Friedmann equations}

The cosmological applications of the hybrid metric-Palatini gravitational theory have been extensively explored \cite{Capozziello:2012ny, Harko:2011nh}, in the scalar-tensor representation. In particular, it was shown that the accelerating expansion and, in general, any cosmological behaviour, strictly depends on the effective scalar field potential. However, it is rather important to stress that the scalar field has a purely geometric origin and describes further degrees of freedom of the gravitational field coming from extended theories of gravity.

Now, consider the Friedmann-Robertson-Walker (FRW) line element given by
\ba
ds^2=-dt^2+a^2(t)\left[ \frac{dr^2}{1-Kr^2} +r^2 (d^2 \theta + \sin^2\theta d^2\phi) \right]\,,
\ea
where $K = +1, 0, −1$ corresponds to a closed, flat, and open universe, respectively.
For this case, the scalar curvature takes the following form
\begin{equation}
R=6\left(2H^2+\dot{H}+ \frac{K}{a^2} \right) \,.
\end{equation}

The effective Friedmann equations can be written in terms of the effective energy density and pressure, given by
\ba
3H^2 & = & \kappa^2\rho_{\rm eff} - \frac{3K}{a^2}\,, 
\label{fr1} \\
\dot{H} & = & -\frac{\kappa^2}{2}\lp\rho_{\rm eff}+p_{\rm eff}\rp + \frac{K}{a^2} 
\label{fr2} \,,
\ea
with the following relationships
\ba
\lp 1+\phi \rp \kappa^2\rho_{\rm eff} & = & -\frac{3}{4\phi}\dot{\phi}^2 + \frac{1}{2}V(\phi) - 3H\dot{\phi} + \kappa^2\rho_m\,, 
\label{m1}\\
\lp 1+\phi \rp \kappa^2 p_{\rm eff} & = & -\frac{3}{4\phi}\dot{\phi}^2 - \frac{1}{2}V(\phi) + \ddot{\phi} + 2H\dot{\phi} + \kappa^2 p_m\,.
  \nonumber \\
\label{m2}
\ea  

The conservation equations for the matter component and the scalar field are
\ba
\dot{\rho}_m + 3H(\rho_m + p_m) & = & 0\,, 
\label{ma}\\
\ddot{\phi} + 3H\dot{\phi} - \frac{\dot{\phi}^2}{2\phi} + \frac{1}{3}\phi R - \frac{1}{3} \phi V'(\phi) & = & 0  
\label{kg}\,.
\ea

\subsection{Einstein static Universe}

For the Einstein static universe we choose $a = a_0 = {\rm const}$, so that $H=\dot{H}=0$, and the curvature scalar reduces to $R = 6K/a^2_0$. Note that we do not specify $K$ at this point. We also consider that the matter distribution obeys the following linear equation of state $p_m = w \rho_m$. In classical General Relativity, the presence of a positive cosmological constant allows us to solve the analogue equations of~(\ref{fr1}) and~(\ref{fr2}) under the staticity assumptions with $K=1$. With the above considerations, the effective Friedmann equations (\ref{fr1}) and~(\ref{fr2}) reduce to
\begin{align}
  \kappa^2\rho_{\rm eff} &= \frac{3K}{a_0^2}\,, 
  \label{bg1}\\
  \frac{\kappa^2}{2}\lp\rho_{\rm eff}+p_{\rm eff}\rp &= \frac{K}{a_0^2}\,,
  \label{bg2}
\end{align}
respectively, which imposes the following condition on the distribution of matter
\begin{align}
  \rho_{\rm eff} + 3 p_{\rm eff} = 0\,.
  \label{bg3}
\end{align}

As a first approach, let us now assume that $\phi = \phi_0 = {\rm const}$, then the condition on the effective matter equation implies
\begin{align}
  \kappa^2\rho_m(1+3w) =  V(\phi_0) \,,
  \label{bg4}
\end{align}
and the modified Klein-Gordon equation (\ref{kg}) reduces to the following expression
\begin{align}
  \frac{6K}{a^2_0} = V'(\phi_0) \,.
  \label{bg5}
\end{align}
Therefore, we can express $\rho_m$ and $a_0$ in terms of $\phi_0$ and the potential $V$. Substituting back into one of the two field equations, we can find an implicit equation which fixes the value of the field $\phi_0$ by
\begin{align}
  1+ \phi_0 = \frac{V(\phi_0)}{V'(\phi_0)}\frac{3(1 + w)}{(1+3w)} \,.
  \label{bg6}
\end{align}
Depending on the functional form of the potential, it may be possible to solve for $\phi_0$ explicitly. 

One should also note at this point that $K=-1$ is a possible parameter for an Einstein static universe, in the context of the hybrid metric-Palatini gravitational theory. Provided $V'(\phi_0) < 0$, one can find the scale factor $a_0$ using Eq.~(\ref{bg5}).

\section{Scalar perturbations}\label{sect:4}

It is interesting that the Einstein static universe has been revived as the asymptotic origin of an emergent universe, more specifically as an inflationary cosmology without a singularity \cite{Ellis:2002we}. Despite the fact the positive curvature is negligible at late times, its role is crucial during the early universe. In the latter context, it allows these cosmologies to inflate and later reheat to a hot big-bang epoch. In fact, these cosmological models possess attractive features such as the absence of a singularity, of an `initial time', of the horizon problem, and the quantum regime can even be avoided.

To study perturbations of the Einstein static universe, we will follow the conventions and notation of~\cite{Seahra:2009ft}. We will work in the longitudinal gauge so that the perturbed metric with scalar perturbations is given by
\begin{align}
  ds^2 = a^2(\eta)\left[-(1-2\Psi)d\eta^2 + (1+2\Phi)\gamma_{ij}d\theta^id\theta^j\right]\,
  \label{ip1}
\end{align}
where $i,j = 1,2,3$ and $\gamma_{ij}$ is the metric of the constant curvature 3-space is given by
\begin{align}
  \gamma_{ij} d\theta^i d\theta^j = 
  \frac{dx^2+ dy^2 +dz^2}{\left(1+\frac{k}{4}(x^2+y^2+z^2)\right)^{2}}\,.
\end{align}
We are choosing to work with Cartesian coordinates so that $\theta^1 = x,\ldots$. Note that we cannot restrict ourselves to positive curvature as the background equations also allow for an Einstein static universe with negative curvature.

Recall that the energy-momentum tensor for the perfect fluid can be written
\begin{align}
  T_{\mu\nu} = (\rho + p)u_\mu u_\nu + p g_{\mu\nu}\,,
\end{align}
where the fluid's 4-velocity is given by $u^\mu = (1/a) \delta^\mu_t$. The perturbed matter is given by
\begin{align}
  \delta T^\mu{}_{\nu} = \delta \rho u^\mu u_\nu + 
  u^\mu D_{\nu} v + u_\nu D^{\mu} v + \delta p P^\mu{}_\nu\,.
  \label{ip2}
\end{align}
Here, $v$ is the velocity perturbation and $P^\mu{}_\nu$ and $D_{\nu}$ are given by
\begin{align}
  P^\mu{}_\nu &= \delta^\mu_\nu + u^\mu u_\nu\,, \\
  D_{\nu} &= P^\sigma{}_\nu \partial_\sigma = \partial_\nu + u_\nu u^\sigma \partial_\sigma\,.
\end{align}
Let us also define the currents
\begin{align}
  J_{\nu}^{\rm (eff)} &= \nabla_{\mu} T^{\rm (eff)}{}^{\mu}{}_{\nu}\,, 
  \label{ip_j1}\\
  J_{\nu}^{\rm (m)} &= \nabla_{\mu} T^{\rm (m)}{}^{\mu}{}_{\nu}\,.
  \label{ip_j2}
\end{align}
Finally, we will write the perturbed scalar field as
\begin{align}
  \phi = \phi_0(\eta) + \delta \phi(\eta,\theta^i)\,.
  \label{ip3}
\end{align}

Recall that the Einstein static universe is characterised by the condition that all field values are constant. As in the previous sections, this will be denoted by a subscript 0. Next, we will compute the full Einstein field equations using~(\ref{ip1})--(\ref{ip3}) and linearize with respect to the perturbations. Since these calculations are quite lengthy, we only state the necessary results. In the background, the off-diagonal field equations are identically satisfied, however, in first order perturbation theory, the off-diagonal are non-zero. The three $(0i)$ field equations are given by
\begin{align}
  -2 \partial_i \Phi' = 
  \frac{1}{(1+\phi_0)}\partial_i \partial_\eta \delta\phi -
  \frac{a_0 \kappa^2}{(1+\phi_0)} \partial_i v\,,
\end{align}
where the prime stands for differentiation with respect to conformal time $\eta$. Integration with respect to $\theta^i$ gives
\begin{align}
   -2 \Phi' = 
  \frac{1}{(1+\phi_0)} \delta\phi' -
  \frac{a_0 \kappa^2}{(1+\phi_0)} v\,,
  \label{neq1}
\end{align}
which means that the velocity perturbation $v$ is determined by the perturbation of the spatial curvature $\Phi$ and the scalar field perturbation $\delta\phi$. The three independent $(ij)$ $i \neq j$ field equations read
\begin{align}
  \frac{K/2}{1+k r^2/4}\left[\theta_i \partial_j(\Psi-\Phi) + 
    \theta_j \partial_i(\Psi-\Phi)\right] + \partial_{ij}(\Psi-\Phi) 
  \nonumber \\
  = \frac{1}{(1+\phi_0)}\left\{\frac{K/2}{1+\frac{K}{4}r^2}\left[
    \theta_i \partial_j \delta\phi + \theta_j \partial_i \delta\phi\right] + 
  \partial_{ij} \delta\phi\right\}\,,
\end{align}
which one can solve immediately by noting that
\begin{align}
  \Psi-\Phi = \frac{1}{(1+\phi_0)} \delta\phi\,,
  \label{neq2}
\end{align}
solves the equation. So far we have solved the 6 off-diagonal equations and now address the conservation equations~(\ref{ip_j1}) and~(\ref{ip_j2}). We start by considering the three $J_{i}^{\rm (m)} = 0$ equations which read
\begin{align}
  \partial_i \delta p - \rho_0 (1+w) \partial_i \Psi + 
  \frac{1}{a_0} \partial_i v' = 0\,,
\end{align}
and can be integrated to yield
\begin{align}
  \delta p - \rho_0 (1+w) \Psi + \frac{1}{a_0} v' = 0\,.
  \label{neq3}
\end{align}
One verifies that the $J_{i}^{\rm (eff)} = 0$ equations are equivalent. Here $\Delta$ is the Laplacian of the constant curvature slice, defined by
\begin{align}
  \Delta f = \frac{1}{\sqrt{\gamma\,}}\partial_i
  \left(\sqrt{\gamma\,}\, \gamma^{ij} \partial_j f\right)\,,
\end{align}
where $\gamma$ is the determinant of $\gamma_{ij}$.

On the other hand, the $J_{0}^{\rm (m)} = 0$ (which is also equivalent to $J_{i}^{\rm (eff)} = 0$) reads
\begin{align}
  \delta\rho' + \rho_0(1+w)(3\Phi'-2\Psi') + \frac{1}{a_0} \Delta v = 0\,.
  \label{neq4}
\end{align}

Now we consider the perturbed $(00)$ field equation
\begin{multline}
  -2 \Delta \Phi - 6K (\Phi + \Psi) = \frac{1}{(1+\phi_0)} 
  \Delta \delta\phi \\ 
  + \frac{a_0^2 \kappa^2}{(1+\phi_0)} \delta \rho -
  \frac{a_0^2 \left[V(\phi_0) + 2\kappa^2 \rho_0 (w+2)\right]}{(1+\phi_0)}\Psi \\
  - \frac{a_0^2 \left[2 \kappa^2 \rho_0 - (1+\phi_0) V'(\phi_0) + V(\phi_0)\right]}{2(1+\phi_0)^2} \delta\phi\,.
  \label{neq5}
\end{multline}
Next, we consider the linear combination $(11)+(22)+(33)$ of the field equations which gives
\begin{multline}
  2\Delta(\Phi-\Psi) - 6\Phi'' = 
  -\frac{2}{(1+\phi_0)}\Delta \delta\phi
  + \frac{3 a_0^2 \kappa^2}{(1+\phi_0)} \delta p \\
  + \frac{3 a_0^2(2w\kappa^2 \rho_0 - V_0)}{(1+\phi_0)} \Phi
  + \frac{3}{(1+\phi_0)} \delta\phi'' \\
  - \frac{3 a_0^2 [2w \kappa^2 \rho_0 - V_0 + (1+\phi_0)V'_0]}{2(1+\phi_0)^2} \delta\phi \,.
  \label{neq6}
\end{multline} 
Lastly, we consider the perturbation of the Klein-Gordon equation~(\ref{eq:evol-phi}), which reads
\begin{align}
  -\Box \delta\phi + 
  \bigl[ (1&-3w)\kappa^2 \rho_0 + 2V_0 
  \nonumber \\
  &- V'_0 - \phi_0(1+\phi_0)V''_0 \bigr]
  \frac{\delta\phi}{3} 
  = \frac{\phi_0 \kappa^2}{3} \delta T\,,
  \label{neq7}
\end{align}
where the perturbation of the trace of the energy-momentum tensor is given by
\begin{align}
  \delta T = -\delta\rho + 3\delta p + 2(1+w)\rho_0 \Psi\,.
\end{align}

We now have 7 equations, namely (\ref{neq1}), (\ref{neq2}), (\ref{neq3}), (\ref{neq4}), (\ref{neq5}), (\ref{neq6}), and (\ref{neq7}) for six unknown quantities which are $\Psi$, $\Phi$, $\delta\rho$, $\delta p$, $v$ and $\delta\phi$. It is well known that in General Relativity the seven analogue equations are not independent, in fact there are only 5 independent equations. In order to close the system of equations, one has to prescribe an equation of state for the perturbed matter. It turns out that the situation is the same here, there are only 5 independent equations. Showing this explicitly is rather lengthy but otherwise straightforward. Thus, we also choose and equation of state of the form $\delta p = w \delta \rho$ which corresponds to considering adiabatic single fluid perturbations. Moreover, we will make the usual decomposition into harmonic functions
\begin{align}
  \Psi = \Psi_n(\eta) Y_n(\theta^i)\,, \quad \Phi = \Phi_n(\eta) Y_n(\theta^i) \,,
\end{align}
with $\Delta Y_n = -\nu^2 Y_n$. For $K=1$ we have $\nu^2 = n(n+2)$ for $n=0,1,2,\ldots$; while for $K=-1$ there holds $\nu > 1$.

We start by solving (\ref{neq2}) for $\delta\phi$ and (\ref{neq1}) for $v$ and eliminate those from the other equations accordingly. Next, Eq.~(\ref{neq5}) is solved by $\delta\rho$ and substituted into the remaining 4 equations. We are now inserting the background solution (\ref{bg4})--(\ref{bg6}) into those equations and we find that we are left with two independent equations, namely (\ref{neq3}) which is equivalent to (\ref{neq6}), and (\ref{neq7}). These two independent equations can be written as
\begin{align}
  \begin{pmatrix}
    \Psi''\\ \Phi''
  \end{pmatrix}
  = \mathbf{D}
  \begin{pmatrix}
    \Psi\\ \Phi
  \end{pmatrix}\,,
  \label{eq:finalpert}
\end{align}
where $\mathbf{D}$ is a $2 \times 2$ constant coefficient matrix whose components are complicated expressions involving the background quantities $V_0,V'_0,V''_0$, the spatial curvature $k$ and the equations of state parameter $w$. 

The matrix $\mathbf{D}$ is defined by 
\begin{align}
 \mathbf{D} =
  \begin{pmatrix}
    D_{11} & D_{12} \\
    D_{21} & D_{22}
  \end{pmatrix}\,,
\end{align}
where the components of $\mathbf{D}$ quite complicated. However, some linear combinations are slightly simpler, such as the following relations
\begin{widetext}
\begin{align}
  D_{11} + D_{21} &= K (2-4w) - w \nu^2\,,\qquad D_{12} + D_{22} = w (6K - \nu^2)\,,
  \nonumber \\
  3 V'_{0}(1+3w)(D_{11}-D_{22}) &= -3 K \bigl[4 V'_{0} w (1+3w) + 3 (w^2-1)\bigr]
  + \bigl[V'_{0} (1-9 w^2) + 3 (w+1)(3 w-1)\bigr]\nu^2\,,
  \nonumber \\
  3 V'_{0}(1+3w)(D_{12}-D_{21}) &= 3 K \bigl[2V'_{0}(9w^2-1) - 3w^2 + 3\bigr]
  + \bigl[V'_{0} (1-9 w^2) + 3 (w+1)(3 w-1)\bigr]\nu^2\,.
\end{align}
\end{widetext}

\subsection{Stability of perturbations}

The linear system of equations~(\ref{eq:finalpert}) consists of two coupled second order ODEs and thus will have four linearly independent solutions. Their behaviour is characterised by the two eigenvalues of $\mathbf{D}$ which we denote by $\lambda_1$ and $\lambda_2$. The solutions to~(\ref{eq:finalpert}) will involve the frequencies $\pm\sqrt{\lambda_1}$ and $\pm\sqrt{\lambda_2}$.

Therefore, in order for perturbations to be stable we must require the following four conditions
\begin{align}
  \Re \lambda_1 < 0\,, \qquad \Re \lambda_2 < 0\,,\\
  \Im \lambda_1 = 0\,, \qquad \Im \lambda_2 = 0\,.
\end{align}

Due to the involved components of $\mathbf{D}$ we cannot arrive at analytical results which satisfy our stability condition. However, it is easy to do this numerically and create plots indicating those regions where the perturbations are stable and unstable. We note that we can always rescale the values of the potential at $\phi_0$, therefore we will set $V(\phi_0)=1$. Therefore, for a fixed value of $w$, the two conditions now depend on the two values $V'(\phi_0)$ and $V''(\phi_0)$ and the number $\nu$. 

\subsection{Homogeneous perturbations}

\begin{figure*}[!ht]
\includegraphics[width=0.45\textwidth]{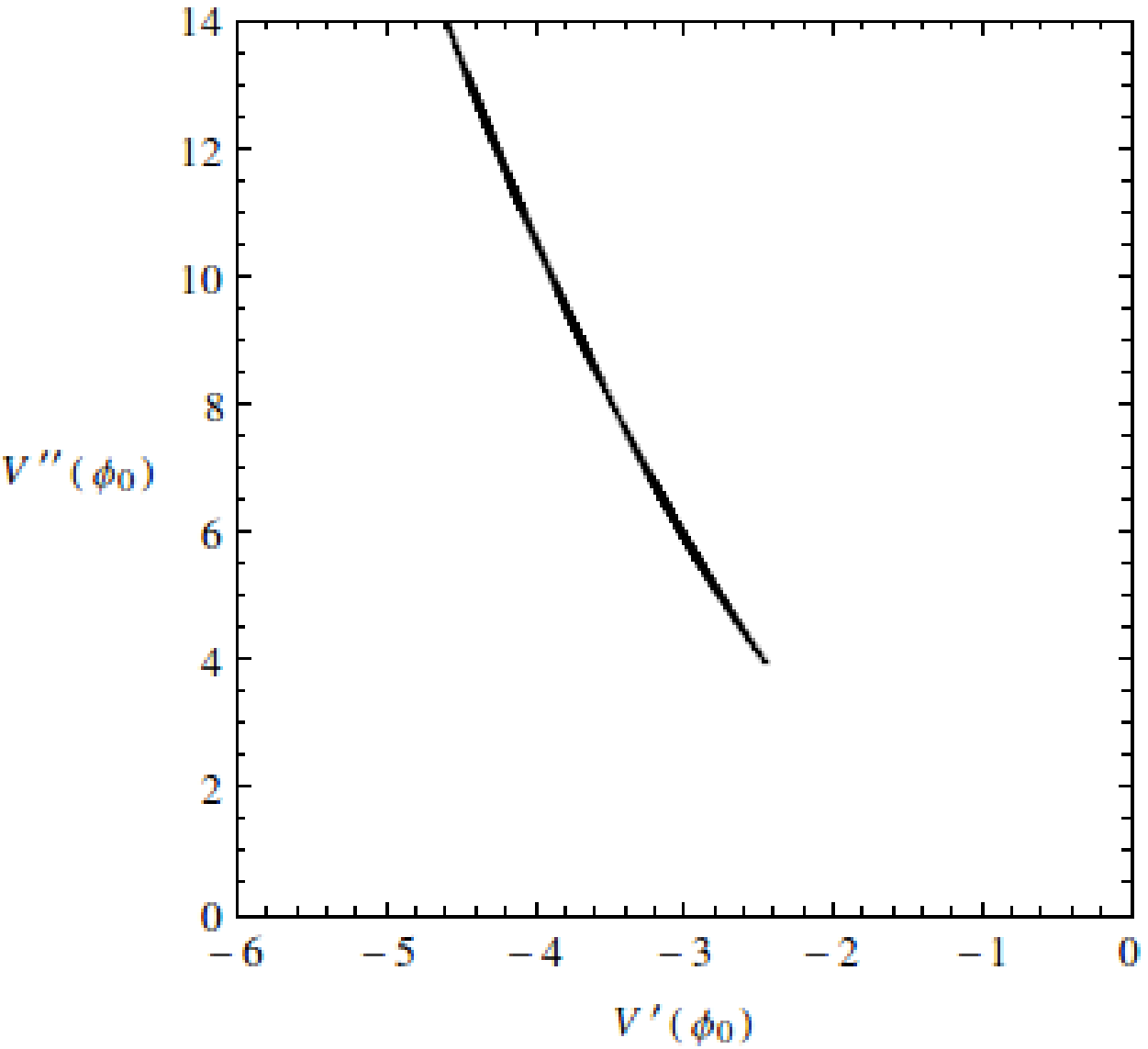}\hfill
\includegraphics[width=0.45\textwidth]{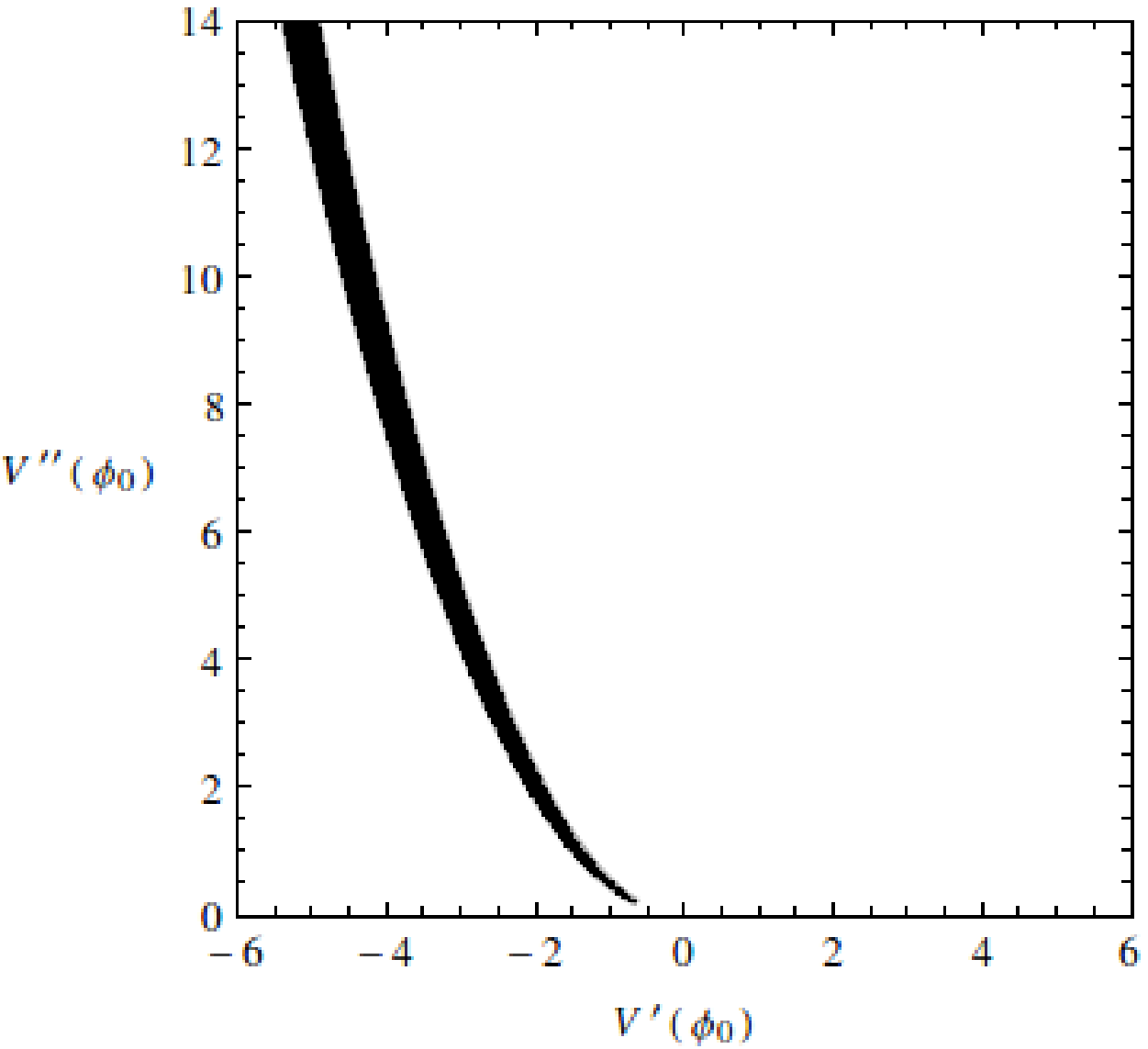}
\caption{Left: Stability regions for $K=1$ and $w=0$. Right: Stability regions for $K=1$ and $w=-1/5$. The shaded regions of the figures represent the parameter space where homogeneous perturbations of the Einstein static universe are stable.} \label{fig:n1}
\end{figure*}

\begin{figure*}[!ht]
\includegraphics[width=0.45\textwidth]{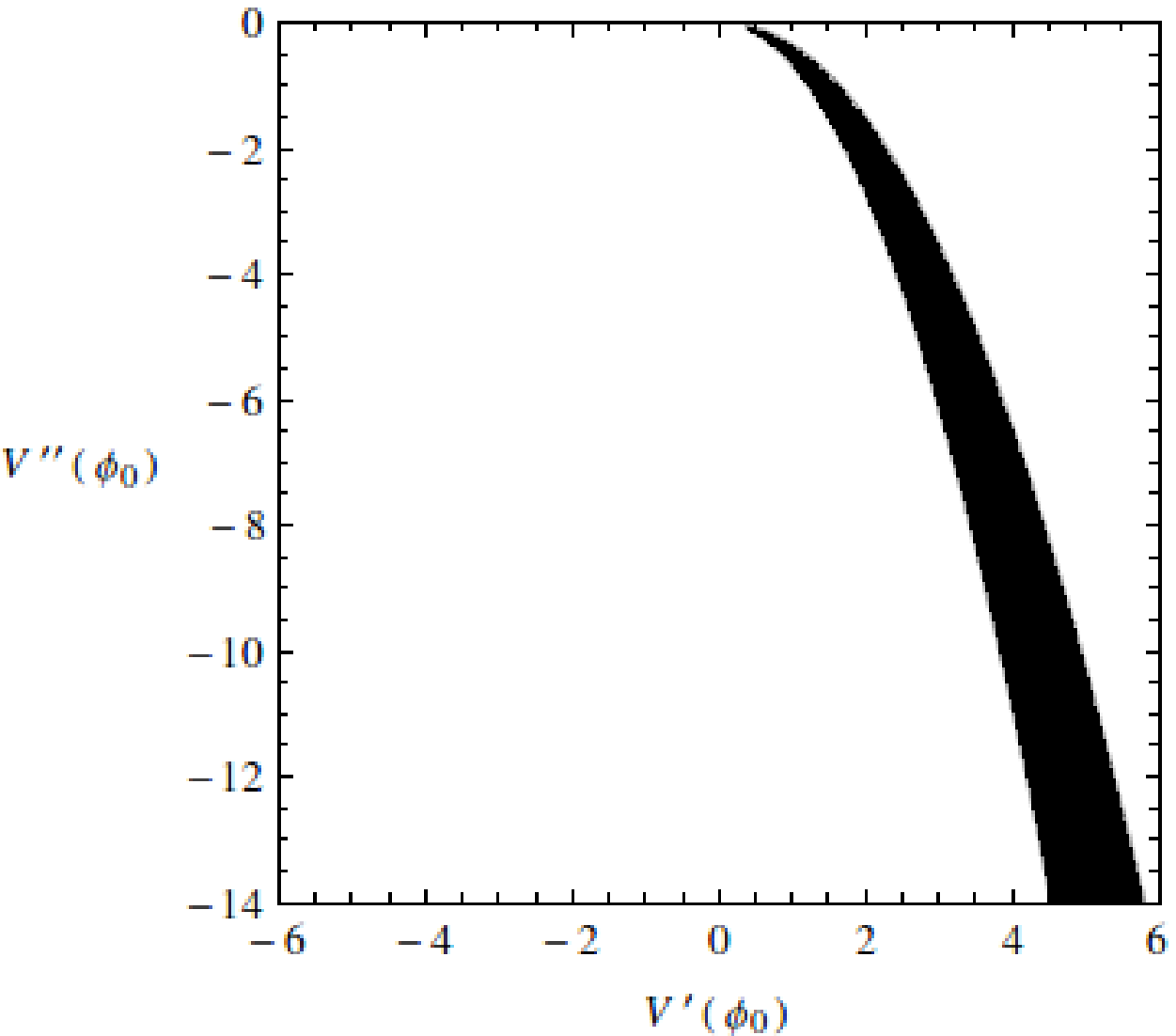}\hfill
\includegraphics[width=0.45\textwidth]{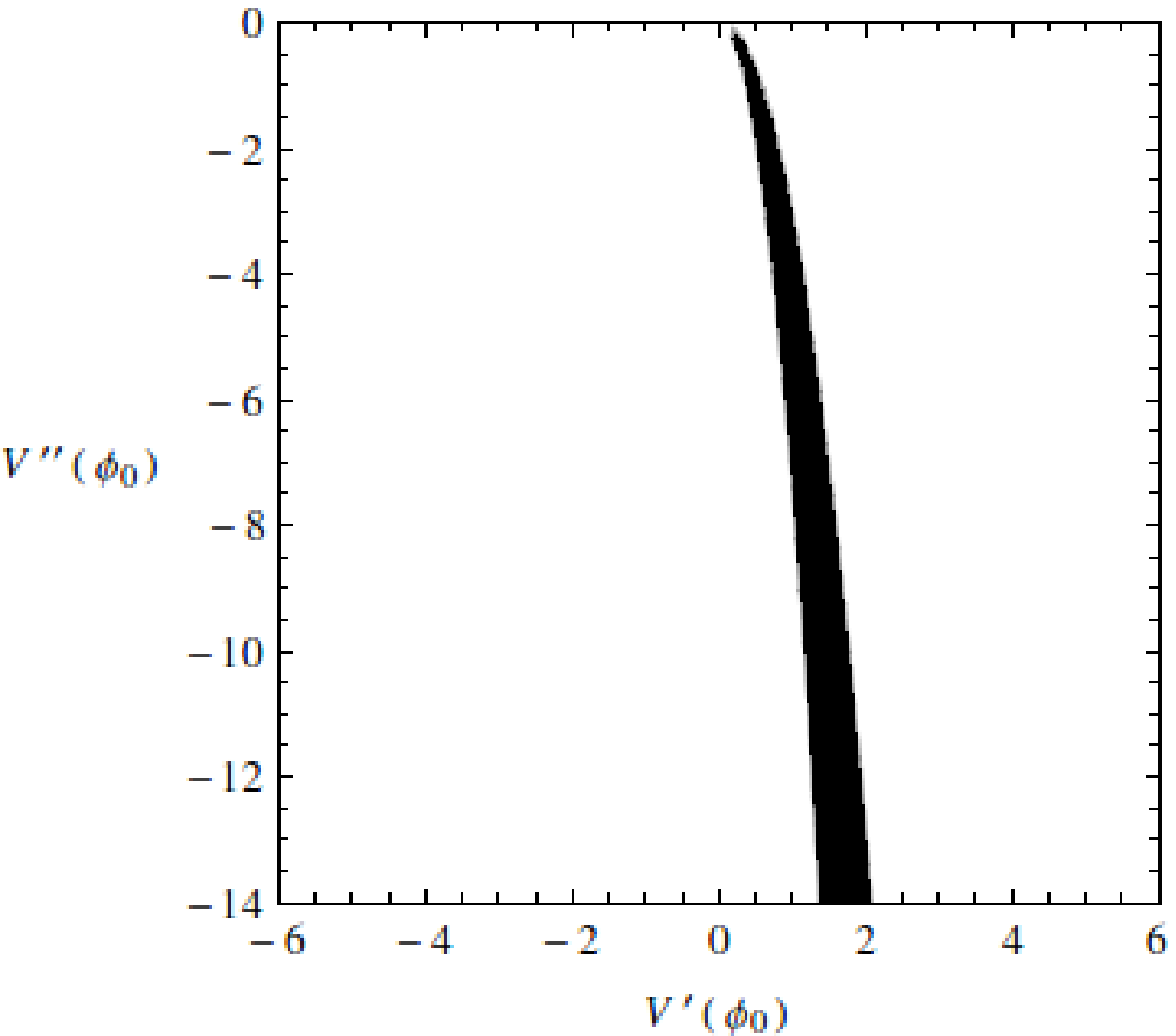}
\caption{Left: Stability regions for $K=1$ and $w=-2/5$. Right: Stability regions for $K=1$ and $w=-3/5$. As in the previous figure, the shaded regions of the figures represent the parameter space where homogeneous perturbations of the Einstein static universe are stable.} \label{fig:n2}
\end{figure*}

When considering homogeneous perturbations, we will set $\nu = 0$ in the matrix $\mathbf{D}$. Note that we can only consider the homogeneous perturbations when $K=1$. For any $w>0$ we could not find any stable solutions. Figures~\ref{fig:n1} and~\ref{fig:n2} show the regions of stability for the equations of state $w = \{0,-1/5,-2/5,-3/5\}$.

We should point out that the parameter spaces for which we find stability might be further reduced when vector and tensor perturbations are taken into account. In this sense, all our stability region are maximal.  


\subsection{Inhomogeneous perturbations}

\begin{figure*}[!ht]
\includegraphics[width=0.45\textwidth]{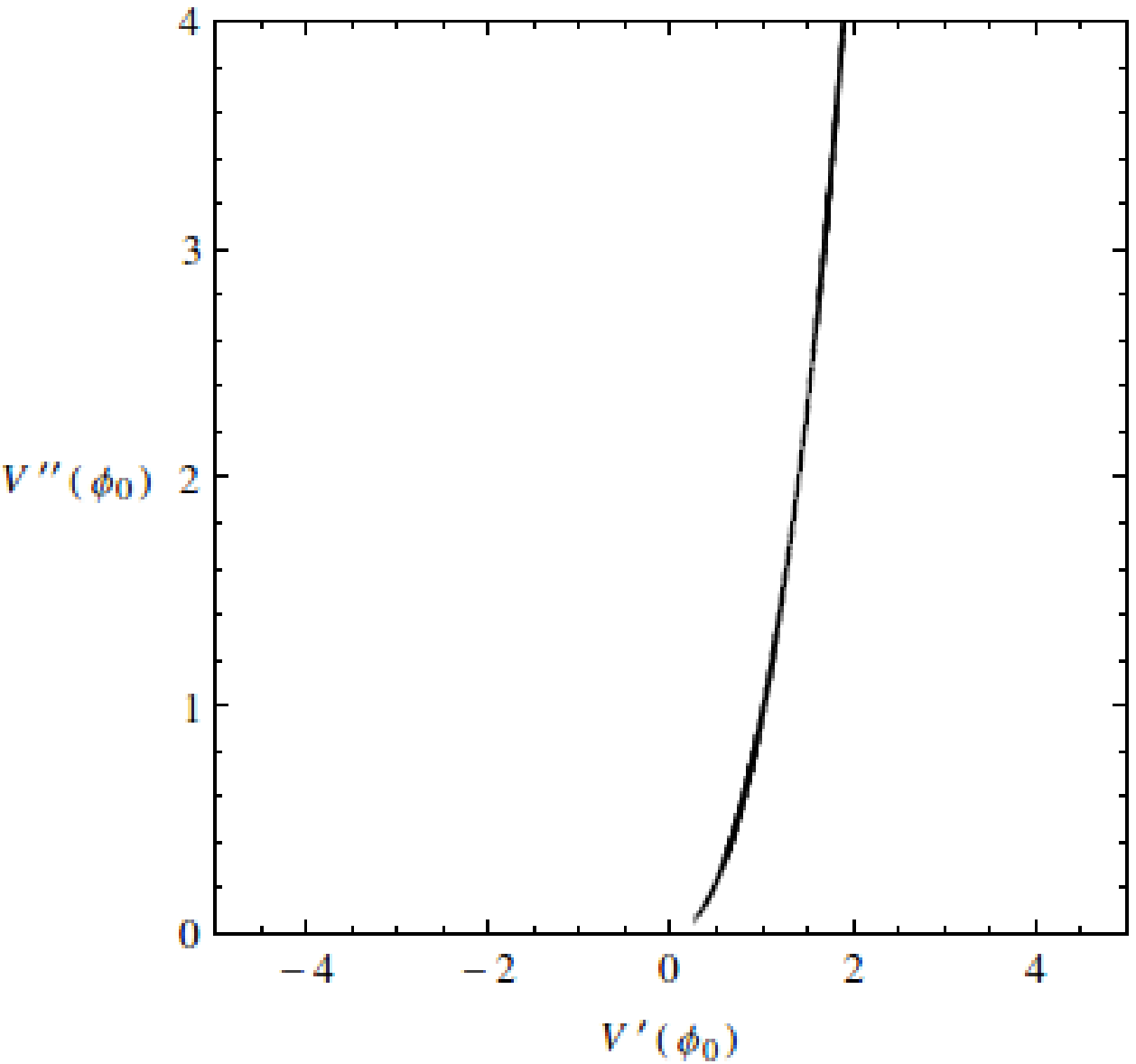}\hfill
\includegraphics[width=0.45\textwidth]{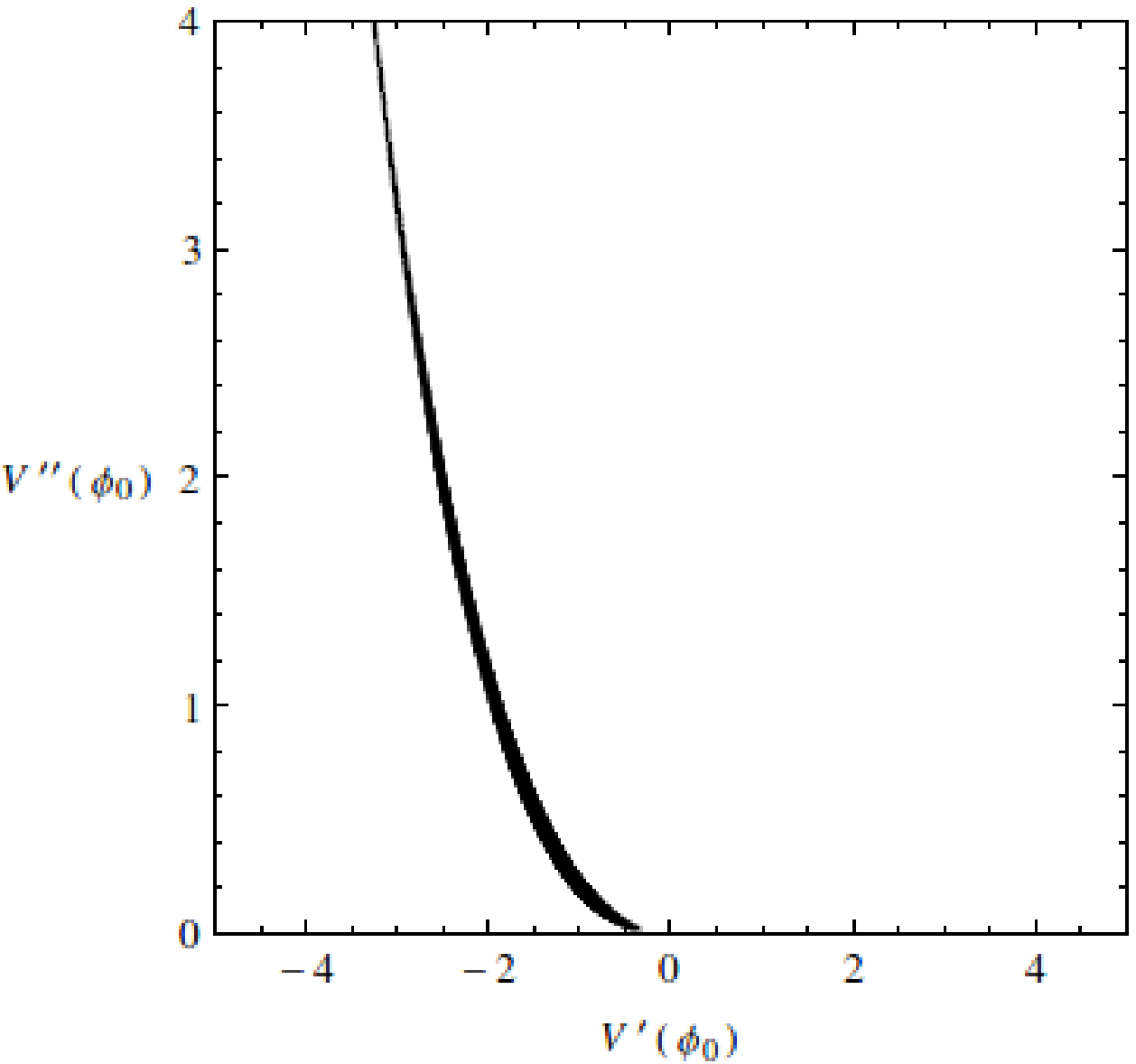}
\caption{Left: Stability regions for $K=1$ and $w=-1/5$. Right: Stability regions for $K=1$ and $w=-2/5$. The shaded regions of the figures represent the parameter space where the $\nu = \sqrt{8}$ or $n=2$ inhomogeneous perturbations of the Einstein static universe are stable.} \label{fig:in1}
\end{figure*}

For $K=1$, we now consider the analogue inhomogeneous perturbations. For concreteness we choose $\nu = 0$ which corresponds to $n=2$. In General Relativity, this corresponds to the largest wavelength perturbation; $n=1$ is a gauge degree of freedom. It turns out that for $w=0$ and $w=-3/5$ we could not identify regions of stability. The remaining two cases are shown in Figure~\ref{fig:in1}.

One notes at once that the regions of stability of the homogeneous and the inhomogeneous perturbations do not overlap. Thus we are led to conclude that we cannot find an Einstein static universe in hybrid metric-Palatini gravity which is stable to both homogeneous and inhomogeneous perturbations, a result very much in chime with the corresponding result in $f(R)$ modified gravity~\cite{Seahra:2009ft}.

On the other hand, if the constant time hypersurfaces are hyperbolic $(K
=-1)$, then there are no homogeneous perturbations. However, we can find solutions to the background equations, see~(\ref{bg5}) and the remark after~(\ref{bg6}). This is in stark contrast to General Relativity where the Einstein static universe cannot exist in this case. Interestingly, we can find stable solutions for $w \geq 0$ but we cannot find those if $w<0$. 

\begin{figure*}[!ht]
\includegraphics[width=0.45\textwidth]{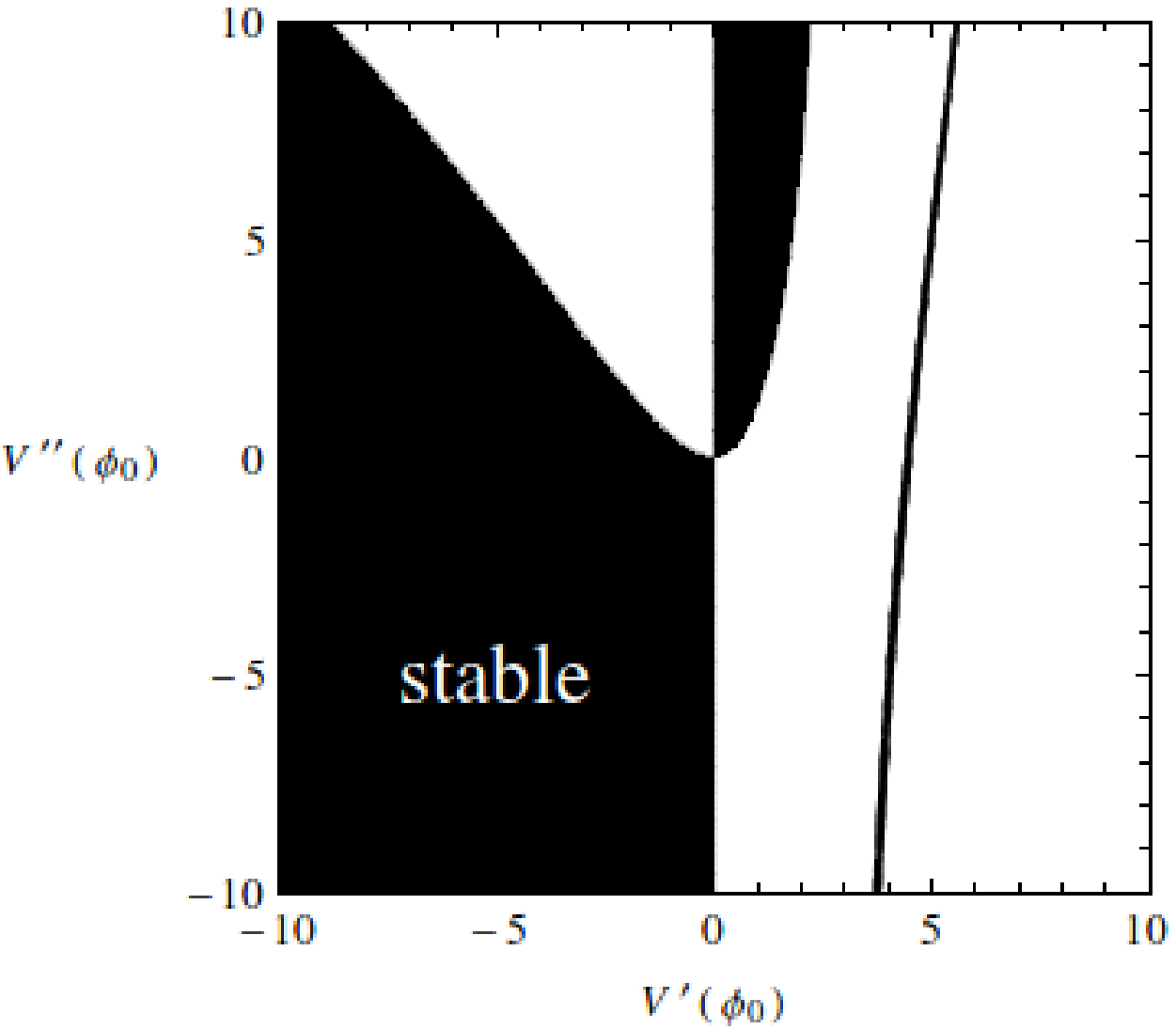}\hfill
\includegraphics[width=0.45\textwidth]{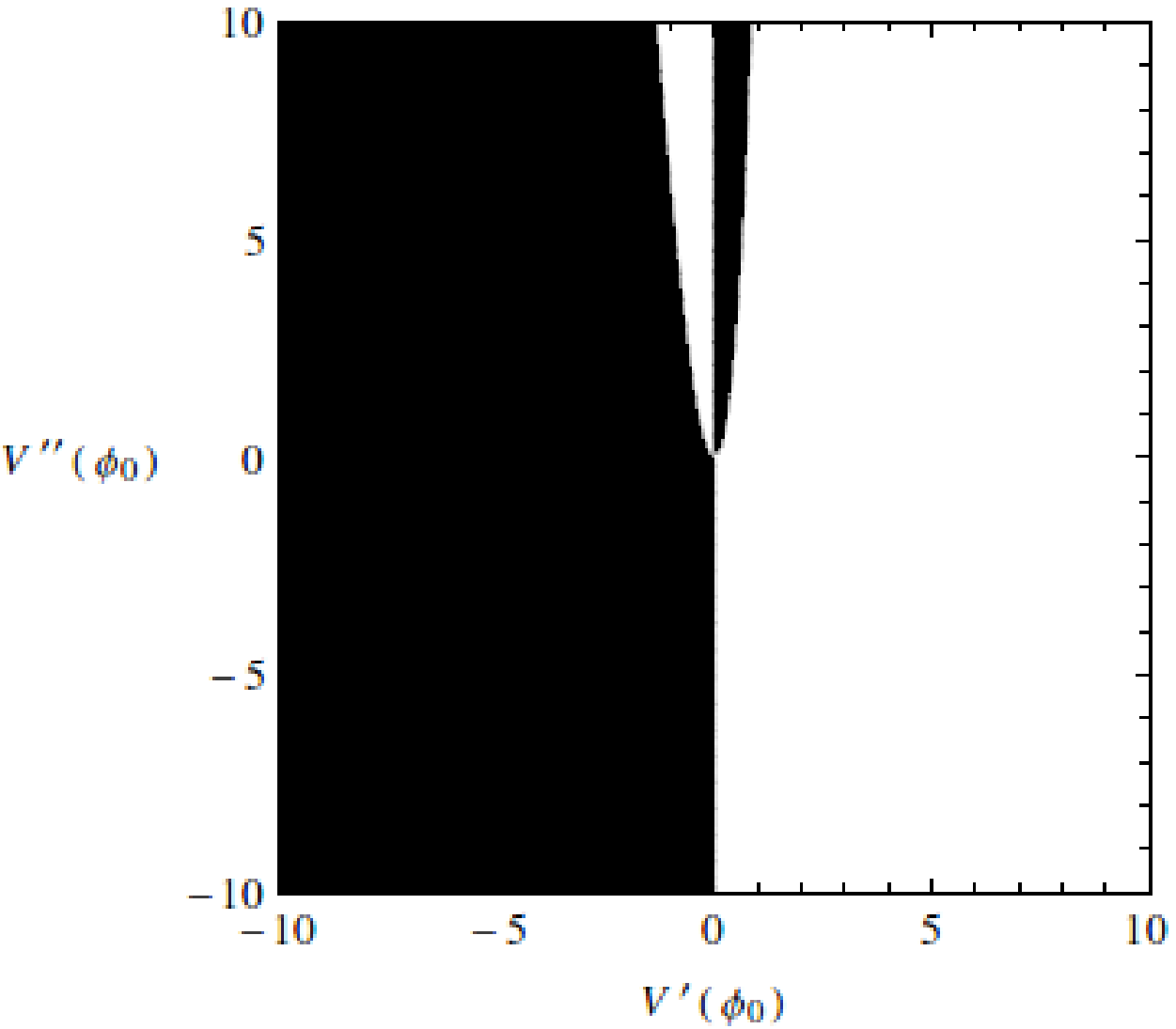}
\caption{Left: Stability regions for $K=-1$, $w=0$ \& $\nu = 1.01$. Right: Stability regions for $K=-1$, $w=0$ \& $\nu = 5$. The shaded regions of the figures represent the parameter space where inhomogeneous perturbations of the Einstein static universe are stable.} \label{fig:km1}
\end{figure*}

\begin{figure*}[!ht]
\includegraphics[width=0.45\textwidth]{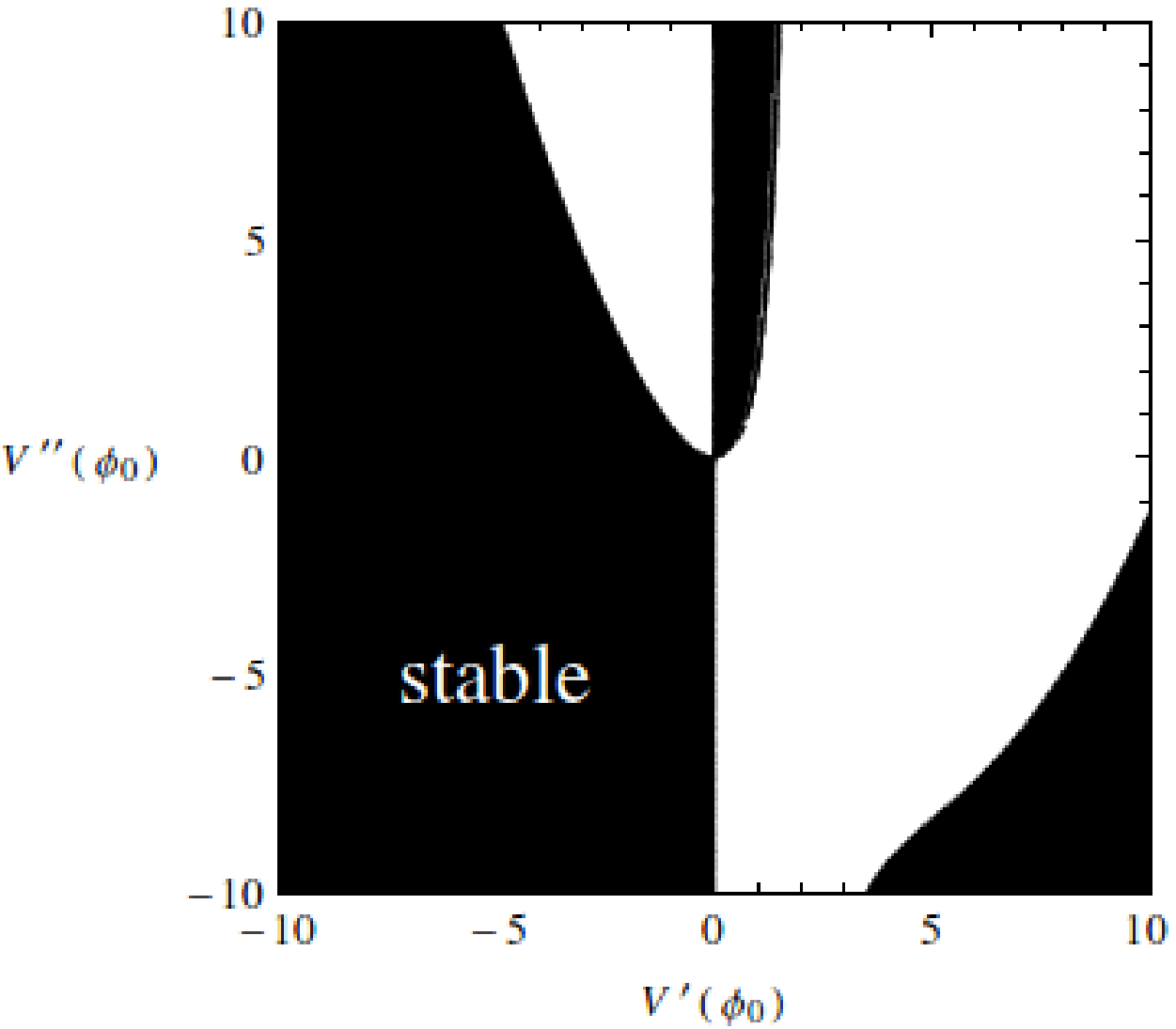}\hfill
\includegraphics[width=0.45\textwidth]{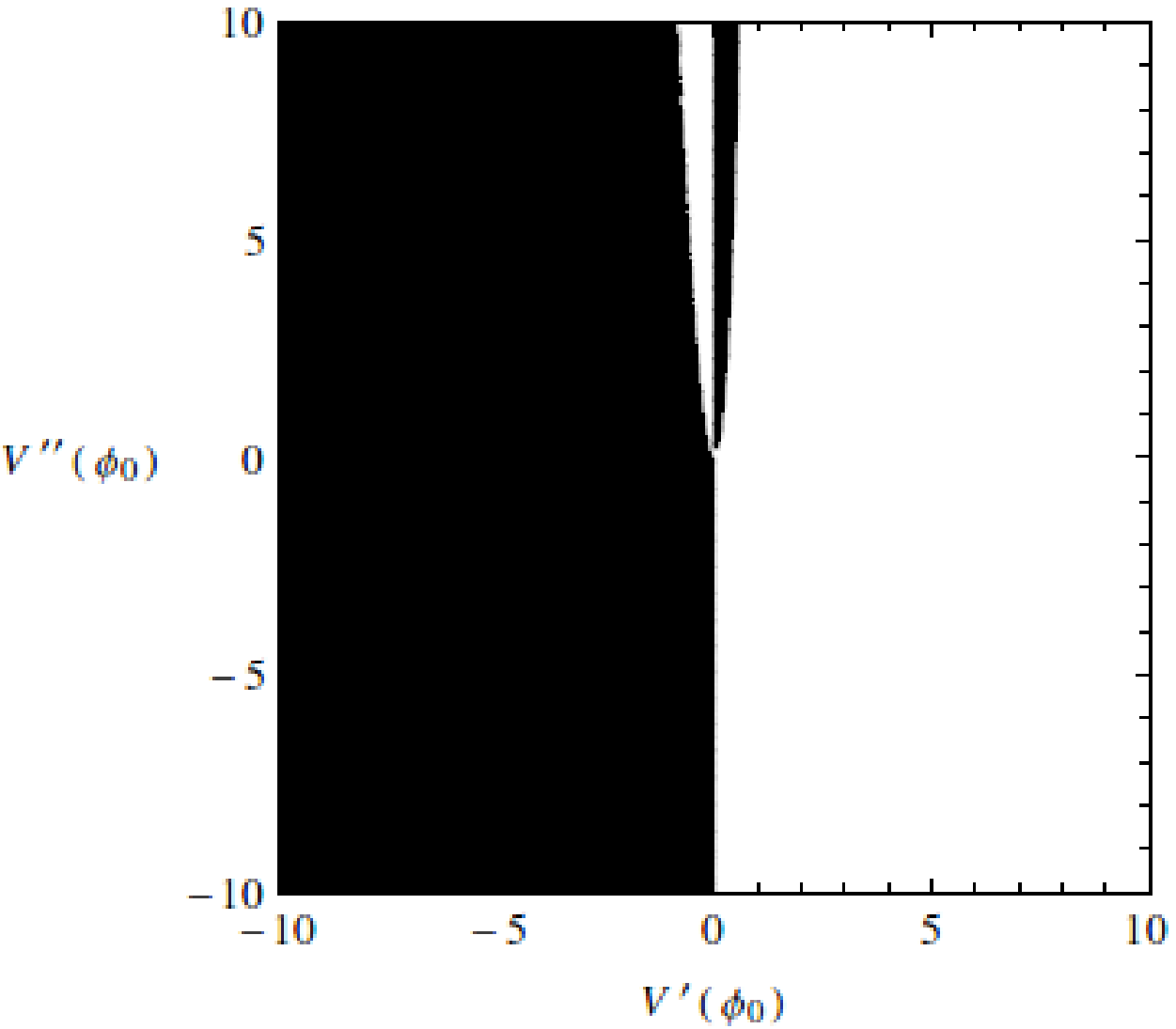}
\caption{Left: Stability regions for $K=-1$, $w=2/5$ \& $\nu = 1.01$. Right: Stability regions for $K=-1$, $w=2/5$ \& $\nu = 5$. The shaded regions of the figures represent the parameter space where inhomogeneous perturbations of the Einstein static universe are stable.} \label{fig:km2}
\end{figure*}

Figures~\ref{fig:km1} and~\ref{fig:km2} show the regions of stability for $w=0$ and $w=1/5$, respectively. In both cases the parabolic shapes at the top of the graphs become narrower as $\nu$ is getting larger and the additional stability region in the $V'(\phi_0) > 0$ region disappears. In either case the region labelled `stable' represent the parameter space for which inhomogeneous perturbations are stable for all $\nu > 1$.  


\begin{figure*}[!ht]
\includegraphics[width=0.45\textwidth]{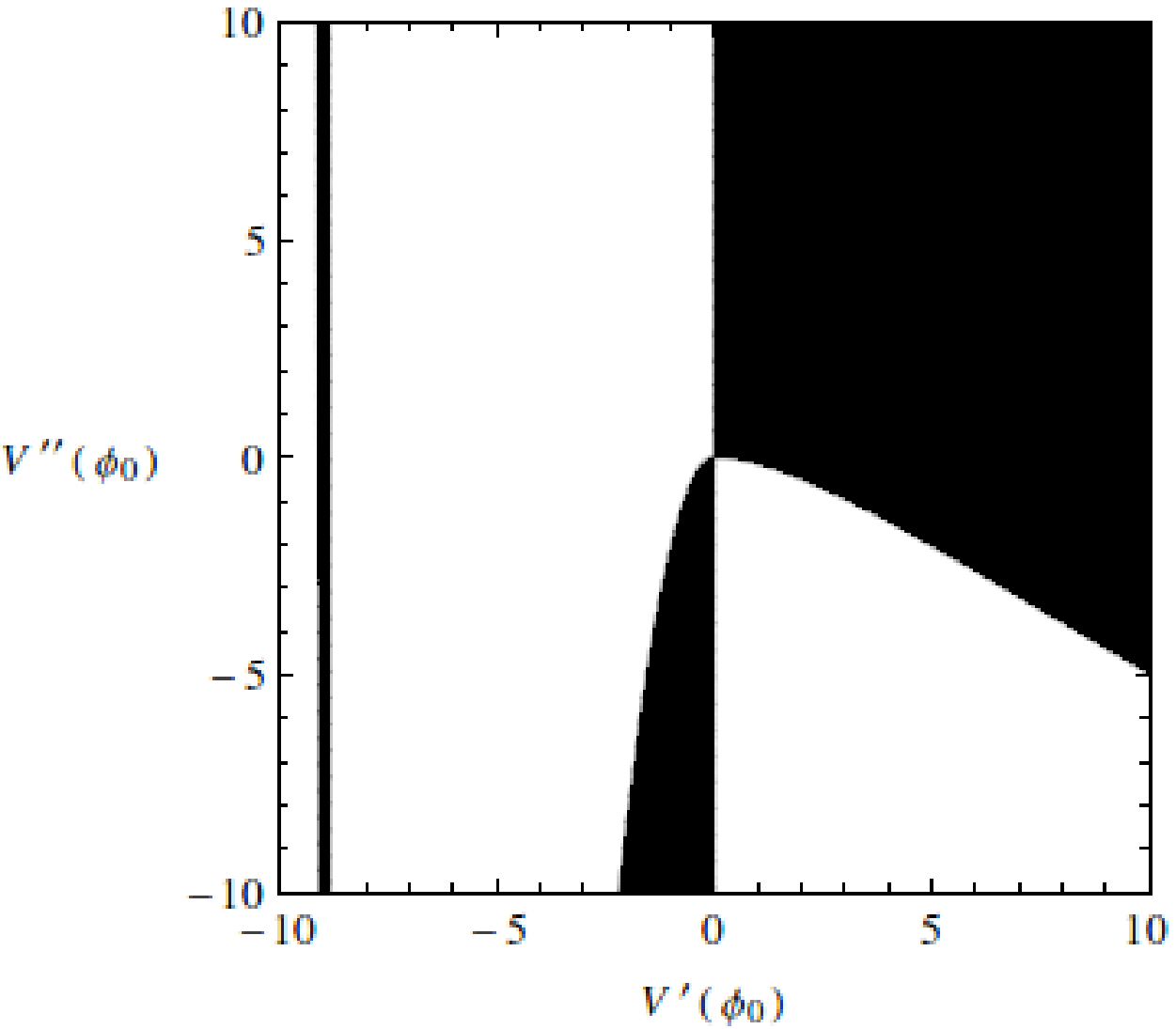}\hfill
\includegraphics[width=0.45\textwidth]{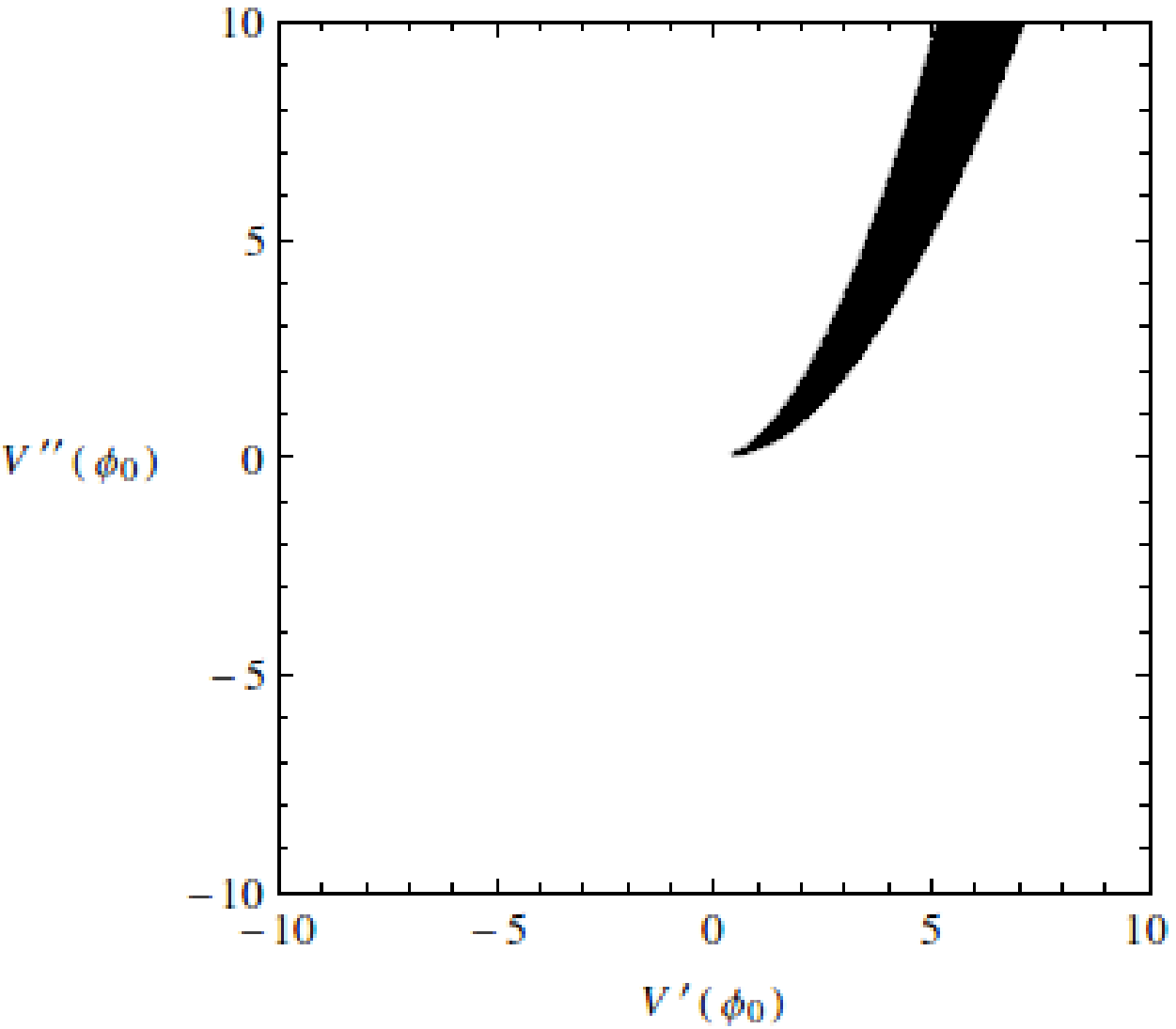}
\caption{Left: Stability regions for $K=-1$, $w=-2/5$ \& $\nu = 1.01$. Right: Stability regions for $K=-1$, $w=-2/5$ \& $\nu = 3$. The shaded regions of the figures represent the parameter space where inhomogeneous perturbations of the Einstein static universe are stable.} \label{fig:km3}
\end{figure*}

This can be compared with Figure~\ref{fig:km3} where we set $w=-2/5$. One notes that the stability region for larger values of $\nu$ becomes increasingly narrower until it will eventually disappears. Thus these solutions would be regarded as unstable.

\section{Summary and Discussion}\label{sec:concl}

In this work, we analysed the stability of the Einstein static Universe by considering linear homogeneous and inhomogeneous perturbations in the respective dynamically equivalent scalar-tensor representation of hybrid metric-Palatini gravity. Considering a linear equation of state parameter for the matter distribution, the stability regions of the Einstein static universe were parametrized by the first and second derivatives of the scalar potential, and it was shown that stable solutions exist in the respective parameter space. The stability regions were considered by requiring positive values for the effective energy density and for the scale factor $a_0$, both of which are independent of the stability of the perturbations.

It is interesting to discuss a few specific choices for the potential $V(\phi)$ and search for stable static solutions. For instance, consider the exponential potential given by
\begin{align}
  V(\phi)=V_0\, e^{-\lambda\phi} \,,
  \label{exppot}
\end{align}
where $V_0$ and $\lambda$ are two positive constants. In an open universe ($K=-1$) a static solution is given by
\begin{align}
  a_0 &= \sqrt{\frac{6}{V_0\,\lambda}}\, \exp\left[-\frac{\lambda +3 (\lambda +1) w+3}{6 w+2}\right] \,,\\
  \rho_0 &= \frac{V_0}{\kappa ^2 (3 w+1)}\, \exp\left[\lambda +\frac{2}{3 w+1}+1\right] \,,\\
  \phi_0 &= -\frac{\lambda +3 (\lambda +1) w+3}{\lambda(1 +3 w)} \,,
\end{align}
where $w>-1/3$ in order to have $\rho_0>0$. This is quite interesting as the effects of hybrid metric-Palatini gravity are sufficiently strong to allow for a static solution in the open case. This solution is stable with respect to inhomogeneous perturbations, provided the parameter $\lambda$ is chosen appropriately, see also Figure~\ref{fig:km1}.

Let us now set $K=1$. Recall that in General Relativity the Einstein static universe only exists if $K=1$. However, for this case, there are no real and positive solutions for $a_0$ in this closed universe with the exponential potential~(\ref{exppot}). This follows from~(\ref{bg5}), the exponential potential has negative derivative and for positive $K$ the scale factor $a_0$ cannot be real. This indicates that one should consider a different potential. 

As a second specific potential, consider the quadratic potential given by
\begin{align}
  V(\phi) = \frac{1}{2} m^2 \phi^2 \,,
\end{align}
where $m$ is a positive constant scalar field mass. A static solution for this potential is given by
\begin{align}
  a_0 &= \sqrt{\frac{3\, K\, (1-3 w)}{A\,(6 w+2)}} \,, \\
  \rho_0 &= \frac{4\, A\, (3 w+1)}{\kappa ^2\, (1-3 w)^2} \,,\\
  \phi_0 &= \frac{6 w+2}{1-3 w} \,.
\end{align}
As in the previous case, the background solution will place constraints on the permissible parameter range. In an open universe ($K=-1$) we must require $w>1/3$ in order to have $a_0$ and $\rho_0$ both real and positive, while in a closed universe ($K=1$) we need $-1/3<w<1/3$. The potential satisfies $V''(\phi) = m^2$ and is thus strictly positive. Combining this with the condition on $w$, we conclude that the open universe solution can be stable under inhomogeneous perturbations. On the other hand, the closed universe is unstable for homogeneous perturbations while the open universe can be stable, see Figures~\ref{fig:km1} and~\ref{fig:km2}.

In conclusion, concentrating the analysis in this work to the stability of the Einstein static Universe by considering linear homogeneous and inhomogeneous perturbations, due to the additional degrees of freedom in the hybrid 
metric-Palatini theory, these lead to enhanced regions of stability in the parameter space. Thus, we have found interesting results that present a richer stability/instability structure that in General Relativity.

\section*{Acknowledgments}
FSNL acknowledges financial support of the Funda\c{c}\~{a}o para a Ci\^{e}ncia e Tecnologia through the grants CERN/FP/123615/2011 and CERN/FP/123618/2011.

\end{document}